\documentclass[12pt]{report}
\renewcommand{\big}{}
\usepackage{graphicx,epsf}
\usepackage{float}
\begin{document}
\topmargin -.4in
\textheight 9.2in
\textwidth 5.5in
\sloppy
\baselineskip .25in
\begin{center}
{\bf Single hole doped strongly correlated ladder with a static impurity }\\

{\bf S. Gayen}\\
Department of Physics, Drexel University, Philadelphia, PA 19104,USA
\end{center}
\vskip .4in
\begin{center}
{\bf Abstract}
\end{center}

\vskip .3 in
\noindent
We consider a strongly correlated ladder with diagonal hopping and exchange 
interactions described by $t-J$ type hamiltonian. We study the dynamics of a 
single hole in this model in the 
presence of a static non-magnetic (or magnetic) impurity. In the case of a 
non-magnetic (NM) impurity we solve the problem analytically both in the 
triplet (S=1) and singlet (S=0) sectors. In the triplet sector the hole doesn't form any bound state with the impurity. However, in the singlet sector the hole forms
bound states of different symmetries with increasing $J/t$ values. Binding 
energies of those impurity-hole bound states are compared with the binding 
energy of a pair of holes in absence of any impurity. In the case of magnetic 
impurity the analytical eigenvalue equations are solved for a large (50 X 2) 
lattice. In this case also, with increasing $J/t$ values,  impurity-hole bound 
states of different symmetries are obtained. Binding of the hole with the 
impurity is favoured for the case of a ferromagnetic (FM) impurity than in the 
case of antiferromagnetic (AFM) impurity. However binding energy is found to be
maximum for the NM impurity. Comparison of binding energies and various 
impurity-hole correlation functions indicates a pair breaking mechanism by 
NM impurity. 
\newpage
\section*{1. Introduction}
Since the discovery of the high-$T_c$ superconductivity \cite{Bednorz} 
the nature of the order parameter, the symmetry of the bound 
pair of holes has been  a hotly debated issue [2-10]. It is well known that 
in the case of conventional superconductors the introduction of magnetic impurities 
drastically suppresses the transition temperature $T_c$. Also theoretically it 
was predicted that magnetic impurities are strong pair 
breakers for conventional superconductors \cite{Abrikosov}, while NM impurities have a pair 
breaking effect only for higher orbital momentum states \cite{Maki, Ueda} such as d-wave pairing 
state. There were extensive studies on the effect of doping the high $T_c$ 
cuprate systems with static magnetic and NM impurities in the last 
decade [14-17]. Experimental studies have shown that divalent Zn and Ni ions replace 
the Cu$^{++}$ ions in the CuO$_2$ planes and ZN (the NM impurity) is 
more effective than Ni (the magnetic impurity) in destroying 
superconductivity. It was suggested by some groups 
\cite{Xiao1,Borkowski,Mahajan} that NM impurities do not simply act as 
vacancies but induce local magnetic moments in the CuO$_2$ planes and the 
mechanism is still a magnetic pair breaking mechanism. However, other 
experimental studies \cite{Walstedt} have indicated that the estimated impurity moment-carrier 
exchange in Zn doped YBCO is too small to account for the suppression of $T_c$
by a magnetic pair breaking mechanism satisfactorily. Moreover, the 
observed difference in behaviour of Zn and Ni doped systems could not be 
properly explained.

In order to explain all these features qualitatively Poilblanc $et al.$ 
initiated numerical studies on a microscopic model, the $t-J$ model. In a series of papers \cite{Poilblanc1,Poilblanc2, Poilblanc3} they studied the 
effect of magnetic and NM impurities in very small clusters  in 2D.  Based upon
their calculation of local density of states in presence of an impurity (NM or magnetic), they found out the existence of successive impurity-hole bound states
of d,s and p-wave symmetries with increasing 
$J/t$ values \cite{Poilblanc1,Poilblanc2}. From the analysis of impurity-hole 
binding energies and quasiparticle weight of a pair of holes \cite{Poilblanc3} they concluded that NM impurity has a stronger pair breaking effect than a magnetic impurity. 
In this respect we also looked at the effect
 of magnetic and NM impurities on single hole dynamics in a $t-J$ ladder
 model introduced by Bose and Gayen \cite{Bose1}. Existence of bound hole pair and superconductiong correlations in a $t-J$ ladder model was 
first numericallly shown by Dagotto $et al.$ in 1992 \cite{Dagotto}. Later Gayen and 
Bose \cite{Gayen, Bose2} proved analytically that the two holes form a bound 
propagating object in the ground state with a modified d-wave symmetry and the 
superconducting correlations exist in a $t-J$ ladder model with diagonal interactions.

We divide the paper in the following sections. In section {\bf 2} we briefly 
introduce the model and recapitulate the exact results in the undoped limit as 
well as for single and two holes doped cases. In section {\bf 3} we derive analytical
 results for single hole dynamics in presence of a static  
NM impurity. In section {\bf 4} we consider the hole in presence of a 
static magnetic impurity.   In the last section (section {\bf 5}) we conclude by speculating the implication of a pair-breaking mechanism by NM impurities in our model.

\section*{2. The Model}
The ladder model shown in figure 1 consists of two chains joined together by rungs and
diagonals. The model is described by the following hamiltonian
\begin{equation}
H=-\sum_{<i,j>\sigma}t_{ij}\big(C_{i,\sigma}^{\dag}C_{j,\sigma} + H.C.\big) +
\sum_{<i,j>}J_{ij}\vec{S_i}\cdot\vec{S_j}
\end{equation}   
The hopping integral along the solid (dashed) line is $t$ ($t^\prime$). The corresponding AFM exchange interaction is $J$ ($J^\prime$). The spins are of magnitude
$\frac{1}{2}$ and single occupancy constraint is strictly enforced in our 
model. We use periodic boundary conditions (PBC). In the undoped limit the ground state consists of singlets along the 
vertical rungs when $2J=J^\prime$ \cite{Bose1} with ground state energy 
$E_G=-\frac{3}{4}J^\prime N$. Actually this is the ground state for 
$J^\prime > 1.40148403897 J $ \cite{Xian}. In this paper we 
shall consider $J^\prime=2J$ for simplicity. However, the results are exact for all values of $J^\prime > 1.40148403897J$.
When the ladder is doped by a single hole, depending upon
the value of $J/t$ the ground state may consist of either the hole bound to an 
adjacent triplet excitation or the hole can coherently propagate through a
sea of singlets \cite{Bose3}. When the ladder is doped with two holes the ground state consists of coherently propagating bound state of two holes of modified d-wave symmetry \cite{Gayen}. 

In this context it is important to note that ladder models are not only of
theoretical interest but also of experimental relevance \cite{Dagotto1}. Hiroi $et al.$ \cite{Hiroi} synthesized the family of compounds 
$Sr_{n-1}Cu_{n+1}O_{2n}$ that 
can be described well by ladder geometry \cite{Rice, Azuma}. M$\ddot{u}$ller and Mikeska
\cite{Muller} discussed about the elementary excitations in the compound $KCuCl_3$ which is described by an AFM ladder model with additional diagonal 
couplings.  The compound $LaCuO_{2.5}$, formed by an array of weakly 
interacting ladders \cite{Hiroi1}, could be doped with holes \cite{Hiroi2} on 
replacing $La$ by $Sr$. Furthermore, the discovery of superconductivity in 
doped ladder compound $Sr_{14-x}Ca_{x}Cu_{24}O_{41}$ by Uehara $et al.$ \cite{Uehara} gave a tremendous impetus to theoretical and experimental research works on ladders.
\begin{figure}[H]
\centering
\includegraphics[width=12cm,height=2.8cm]{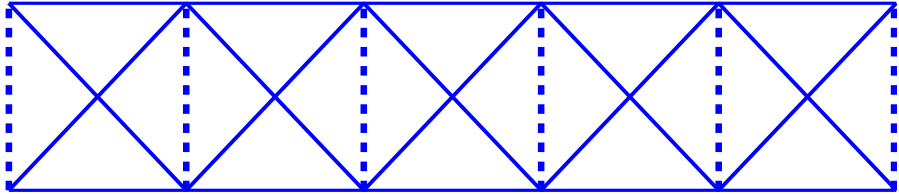}
\caption{The ladder model with diagonal coupling.}
\end{figure}

\section*{3. Non-magnetic Impurity}
The NM impurity is modelled as an inert site located on the first rung of the ladder. PBC is implied. The ground state consists of singlets along all the rungs except  the first rung where there is a free spin-$\frac{1}{2}$ along with the impurity. For $N$ rungs the ground state energy will be $E_g=-\frac{3J^\prime}{4}(N-1)$. A single hole
can be added in the singlet state by removing the free spin. It can be added 
to another rung thereby creating another free spin also. The two free spins in the first rung and in the other rung, where a hole had been added, can combine to form either a singlet (S=0) or triplet (S=1) state. 
\subsection*{3.1 triplet (S=1) sector}
In the triplet sector we can generate a closed subspace of basis states $\phi(1), \phi(2), \cdot \cdot \cdot, \phi(N-1)$,  
\begin{eqnarray}
\phi(1)&=&^{\ast}_{\uparrow}\frac{1}{\sqrt 2}(^{\uparrow}_{O}+^{O}_{\uparrow})||\cdot \cdot \cdot\nonumber\\
\phi(2)&=&^{\ast}_{\uparrow}| \frac{1}{\sqrt 2}(^\uparrow_{O} +^{O}_{\uparrow} )||\cdot \cdot \cdot\nonumber\\
\phi(r)&=&^{\ast}_{\uparrow}||\cdot\cdot\frac{1}{\sqrt 2}(^\uparrow_{O}+^{O}_{\uparrow} )||\cdot \cdot \cdot
\end{eqnarray}
where the argument $r$ implies that the hole is on the ($r+1$)-th rung. The vertical solid line along $i$-th rung stands for a singlet ($\frac{1}{\sqrt{2}}(c_{2i-1\uparrow}c_{2i\downarrow}-c_{2i-1\downarrow}c_{2i\uparrow})$).
A general eigenfunction $\Psi=\sum_{i=1}^{N-1}a_i\phi(i)$ with energy 
eigenvalue $E$ satisfes the following eigenvalue equations:
\begin{eqnarray}
(\epsilon-\frac{J}{4})a_1&=&t a_2\nonumber\\
\epsilon a_2&=&t (a_1+a_3)\nonumber\\
\cdot\cdot\cdot\cdot\cdot\cdot&=&\cdot\cdot\cdot\cdot\cdot\cdot\nonumber\\
\epsilon a_{N-2}&=&t (a_{N-1}+a_{N-3})\nonumber\\
(\epsilon-\frac{J}{4})a_{N-1}&=&t a_{N-2}
\end{eqnarray}
where $\epsilon=E-\frac{3 J^\prime}{4} + t^\prime$. The energy eigenvalue $E$ 
is measured with respect to the ground state energy $E_g$.
The above eigenvalue equations can be solved analytically with symmetric ($a_n=cos(k(N/2-n))$) or antisymmetric ($a_n=sin(k(N/2-n))$) eigenfunctions. In the symmetric case the energy eigenvalues are obtained from simultaneous solution of the equations
\begin{eqnarray}
\epsilon&=&2tcos(k)\nonumber\\
\epsilon-\frac{J}{4}&=&t\frac{cos(k(N/2-2))}{cos(k(N/2-1))}
\end{eqnarray}
In the antisymmetric case the corresponding equations are
\begin{eqnarray}
\epsilon&=&2tcos(k)\nonumber\\
\epsilon-\frac{J}{4}&=&t\frac{sin(k(N/2-2))}{sin(k(N/2-1))}
\end{eqnarray}
In the limit $N \rightarrow \infty$ we get a continuum of extended states for real values
 of $k$. When $J > 4t$  we can derive 
antibound state solutions in both the cases for imaginary values of $k$ ($k$ 
replaced by $ik$). The antibound states are localized states lying above the 
continuum of extended states  with energy eigenvalue $E_{ab}=\frac{3J^\prime}{4}-t^\prime+\frac{J}{4}+\frac{4t^2}{J}$. 
However, for any positive value of $J$ the hole does not form a bound state 
with the impurity in the triplet sector.

\subsection*{3.2 singlet (S=0) sector}
The motion of hole in presence of NM impurity were reported in brief
by Bose and Gayen \cite{Bose4} earlier. For the sake of convenience we briefly 
review the salient features here. A basis state $\phi_i$ will have the hole 
located in the $i$-th rung. There will be singlets residing on all $N-2$ rungs except the first 
and $i$-th rungs. The electronic spins on 1st and $i$-th rungs will form a singlet. In the basis
 state $\phi_1$ both the impurity and the hole are on the first rung. A general
eigenfunction will have the form $\Psi=\sum_{i=1}^N a_i\phi_i$. The exact 
amplitude equations derived from the time independent Schrodinger equation with
energy eigenvalue $E$ (measured with respect to the ground state energy $E_g$) are

\begin{eqnarray}
(\epsilon+\frac{3J^\prime}{4}-t^\prime)a_1&=&t\sqrt{2} (a_2+a_N)\nonumber\\
(\epsilon+\frac{3J}{4})a_2&=&-t\sqrt{2} a_1 + t a_3\nonumber\\
\epsilon a_3&=&t(a_2+a_4)\nonumber\\
\cdot\cdot\cdot\cdot\cdot\cdot&=&\cdot\cdot\cdot\cdot\cdot\cdot\nonumber\\
\epsilon a_{N-1}&=&t (a_{N}+a_{N-2})\nonumber\\
(\epsilon+\frac{3J}{4})a_N&=&-t\sqrt{2} a_1 + t a_{N-1}
\end{eqnarray}
where $\epsilon=E-\frac{3J^\prime}{4}+t^\prime$. In this problem also the 
eigenfunctions are symmetric or antisymmetric with respect to reflection about 
the first rung containing the impurity. In both cases for real values of $k$ 
one gets extended states. With appropriate choice of complex values of $k$ one 
can find out bound ($k \rightarrow \pi +ik$) or antibound ($k \rightarrow ik$) states in the 
symmetric sector. In the case of antisymmetric wave function one only finds 
out bound state solutions. In the $N \rightarrow \infty$ limit, we derive the following results: (a)For the symmetric eigenfunctions,
(i) for $J < \frac{2}{3}[\frac{t^\prime}{2}+2t-\sqrt{8t^2+(\frac{t^\prime}{2})^2}]$, there can be no bound state solution; (ii) for $J > \frac{2}{3}[\frac{t^\prime}{2}-2t+\sqrt{8t^2+(\frac{t^\prime}{2})^2}]$, we do not find any antibound state; (iii) for $J > \frac{2}{3}[\frac{t^\prime}{2}+2t+\sqrt{8t^2+(\frac{t^\prime}{2})^2}]$, we always find two bound states.\\ 
(b) In the case of antisymmetric eigenfunctions, we always get one and only one 
bound state provided $J > \frac{4 t N}{3(N-2)}$. In the  $N\rightarrow\infty$ limit, 
the bound state in the antisymmetric sector has energy 
$E=\frac{3J^\prime}{4}-t^\prime-\frac{3J}{4}-\frac{4 t^2}{3 J}$.

We use the following quantitative definition of the impurity-hole binding
energy introduced by Poilblanc $et al.$ \cite{Poilblanc1, Poilblanc3}
\begin{equation}
\Delta_B(1h,1i)=(E(1h,1i)-E(0h,1i))-(E(1h,0i)-E(0h,0i))
\end{equation}
Figure 2 shows the plot of binding energy ($\Delta_B/t$) as a function of 
exchange interaction ($J/t)$ in different symmetry channels. We also compare the binding energy of a pair of holes computed from the results of Gayen and Bose
\cite{Gayen} in the same figure.  
The binding energy of two holes in absence of impurity is defined as
\begin{equation}
\Delta_B(2h,0i)=(E(2h,0i)-E(0h,0i))-2(E(1h,0i)-E(0h,0i))
\end{equation}
where $E(nh,0i)$ is the ground state energy in presence of $n$ holes and $0$ impurity. $E(1h,1i)$ is the lowest energy in presence of one hole and one impurity in the particular symmetry channel.
\begin{figure}[H]
\centering
\includegraphics[width=11cm,height=6.5cm]{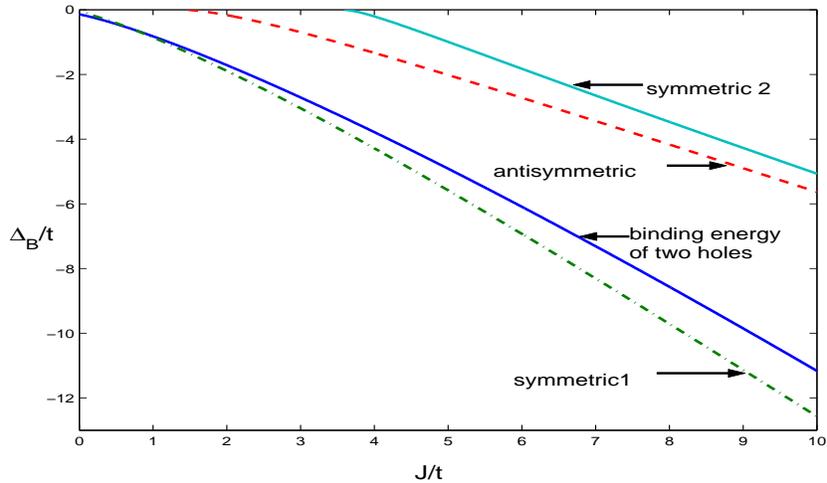}
\caption{Binding energy $\Delta_B/t$ vs. $J/t$ for bound states of different symmetries in the case of NM impurity. The binding energy of a pair of holes is also shown in the figure ($t^\prime/t=1.0$).}
\end{figure}
From figure 2 we find that for any positive value of $J/t$ there is always one bound state in the symmetric (symmetric 1) channel. For $J/t >\frac{4}{3}$ one can find another bound state in the antisymmetric channel. Another bound state is obtained in the symmetric (symmetric 2) channel for all values of $J/t > 3.58152$. The co-existence of the bound states in different symmetry channels are analogous to the appearance of d-, s-, p- wave bound states in 2D clusters studied by
Poilblanc $et al.$ \cite{Poilblanc1}. However, there is one significant 
difference in the ladder. The sequence of the bound states appearing in 
symmetry channels of different parity (even-odd-even) is different from that 
(even-even-odd) obtained by Poilblanc $et al.$ \cite{Poilblanc1}. The most 
significant observation from the graph is that the impurity-hole binding energy in the lowest symmetric sector
has magnitude greater than the binding energy of a pair of holes for $J/t > 0.74$. For 2D finite clusters Poilblanc $et al.$ \cite{Poilblanc1} found the binding energy curve for the pair of holes to be lowest.

\begin{figure}[H]
\centering
\includegraphics[width=11cm,height=6.5cm]{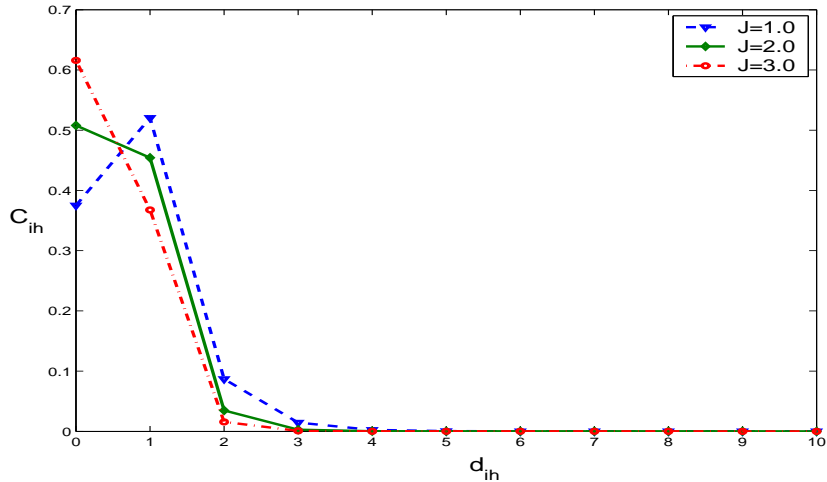}
\caption{Impurity-hole correlation $C_{ih}$ in the ground state as a function 
of the distance $d_{ih}$  of the hole from the NM impurity ($t=t^\prime=1.0$).}
\end{figure}
In order to have more understanding about the nature of impurity-hole binding
we plot (figure 3) the impurity-hole correlation function 
$C_{ih}(j-1)$ ($=<n_i(1)n_h(j)>$) as a function of the distance $d_{ih}$ between 
the impurity and the hole in the ground state. When the hole shares the 
same rung with the impurity, there is no extra cost of energy due to broken 
singlets along the rungs. If the hole sits one rung away from the impurity, 
one additional singlet along the rung is broken. But this loss of energy 
can be compensated by the creation of an extra singlet between the two free 
spins sitting on the two rungs which are occupied by the impurity and hole also.
Moreover, the hole can gain kinetic energy of the amount -$t^\prime$ when it 
stays one rung away from the impurity. As a result the hole likes to sit 
one rung 
away from the impurity in the ground state for smaller values of $J$ 
(with $t=t^\prime=1$).
However, with increasing $J$ values the probability of the hole sharing
the same rung with the impurity increases rapidly.
\section*{4. Magnetic Impurity}
We assume the static impurity to be a spin-$\frac{1}{2}$ object that interacts
with neighbouring spins via exchange interaction $J^\prime_m$ (along the rung) 
and $J_m$ (along horizontal and diagonal bonds). The dynamics of the system is governed by $H_{imp}$ which is essentially the hamiltonian introduced in eqn.(1)
 with the fact that the electrons can not hop on to the impurity site and the 
spin at the impurity site interacts with the neighbouring spins with different 
strengths.
We assume that the static impurity sits on the first rung. We use PBC. In the 
absence of any hole the ground state consists of singlets along all the rungs   with energy 
eigenvalue $-(\frac{3 J^\prime}{4}(N-1)+\frac{3 J^\prime_m}{4})$ if the 
impurity 
exchange interaction is AFM. For FM exchange interaction, all the 
rungs, except the first one, are singlets. The first rung, where the magnetic impurity is located, has $S=1$ configuration. The corresponding energy eigenvalue
is $-(\frac{3 J^\prime}{4}(N-1)+\frac{|J^\prime_m|}{4})$. A single hole can be doped in the first rung to create the state
\begin{equation}
\chi=^{\uparrow^\ast}_{O}||\cdot\cdot\cdot.
\end{equation}
We can diagonalize the hamiltonian within a closed subspace of basis states
denoted by, $\chi$, $\phi(1)$, $\phi(2)$, $\cdot\cdot\cdot$, $\phi(N-1)$,
$\theta(1)$, $\theta(2)$, $\cdot\cdot\cdot$, $\theta(N-1)$,
 $\psi(1)$, $\psi(2)$, $\cdot\cdot\cdot$, $\psi(N-1)$. The states are pictorially represented in the following way
\begin{eqnarray}
\phi(r)&=&(\frac{1}{\sqrt{2}})  ^{\uparrow\ast}_{\uparrow}||\cdot\cdot\cdot(^{O}_{\downarrow}+^{\downarrow}_{O})\cdot\cdot\cdot||\nonumber\\
\theta(r)&=&(\frac{1}{\sqrt{2}}) ^{\uparrow^\ast}_{\downarrow}||\cdot\cdot\cdot(^{O}_{\uparrow}+^{\uparrow}_{O})\cdot\cdot\cdot||\nonumber\\
\psi(r)&=&(\frac{1}{\sqrt{2}}) ^{\downarrow^\ast}_{\uparrow}||\cdot\cdot\cdot(^{O}_{\uparrow}+^{\uparrow}_{O})\cdot\cdot\cdot||
\end{eqnarray} 
where the argument $r$ implies that the  hole is located in the 
($r+1$)-th rung. A general eigenfunction will have the following form 
\begin{equation}
\Psi=a_0 \chi+\sum_{r=1}^{N-1}b_r\phi(r)+\sum_{r=1}^{N-1}c_r\theta(r)+\sum_{r=1}^{N-1}d_r\psi(r)
\end{equation}
Clearly the total number of basis states will be 
$mm=1+3(N-1)$ and we have to diagonalize $mm \times mm$ hamiltonian to find out
 the eigenvalues and eigenfunctions. In our numerical calculations we take $J^\prime_m=2 J_m$, in accordance with the interactions ($J^\prime=2J$) in between the host spins. In this case also the eigenfunctions are found to be symmetric ($b_n=b_{N-n}, c_n=c_{N-n}, d_n=d_{N-n}$) or antisymmetric
($b_n=-b_{N-n}, c_n=-c_{N-n}, d_n=-d_{N-n}$). The lowest energy eigenstate is 
always obtained in the symmetric sector for both FM ar AFM impurity. 
In figure 4 we compare the binding energy of bound states in different symmetry
channels for FM, AFM and NM impurities.
The results obtained indicate that the binding of a doped hole with the impurity is favoured in the case of FM impurity. Binding occurs for weakly AFM impurity also. However, the binding energy in a particular symmetry channel is maximum if it is a NM impurity. In figure 5 we plot the binding energy in the ground 
states for FM ($J_m=-0.5$) and AFM ($J_m=0.5$) impurity and compare them with that in the case of NM impurity.

\begin{figure}[H]
\centering
\includegraphics[width=11cm,height=6.5cm]{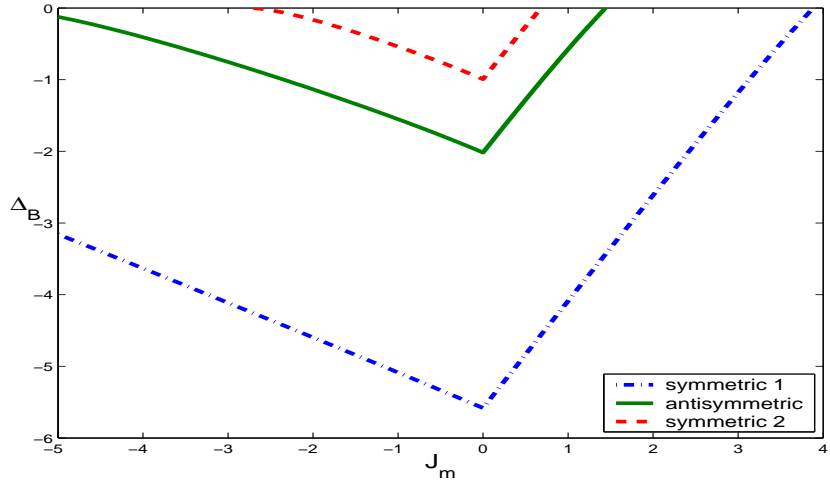}
\caption{Binding energy of bound states of different symmetries as a function of impurity exchange coupling $J_m$ ($J=5.0$,$t=t^\prime=1.0$, $N=50$).}
\end{figure}

\begin{figure}[H]
\centering
\includegraphics[width=11cm,height=6.5cm]{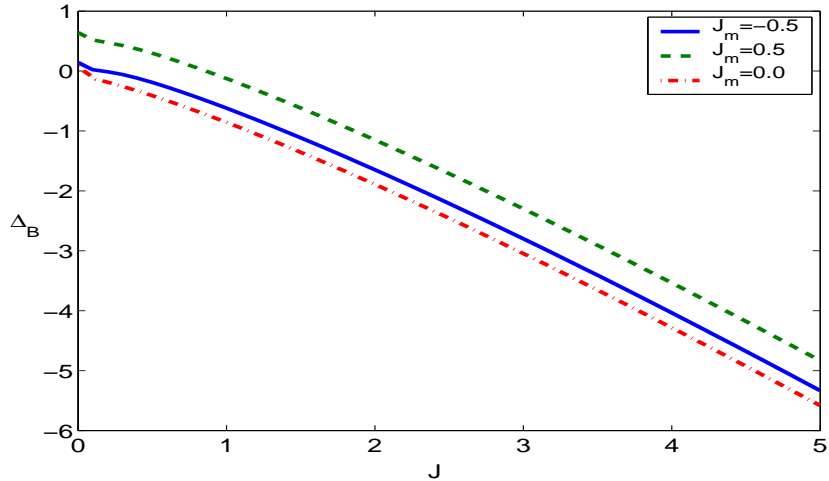}
\caption{Binding anergy of the ground state as a function of $J/t$ for FM ($J_m=-0.5$), AFM ($J_m=0.5$) and NM ($J_m=0.0$) impurity ($t=t^\prime=1.0$, $N=50$).}
\end{figure}

In figure 6 we plot the impurity-hole correlation function ($C_{ih}$) for 
different types of impurities. We notice that the impurity-hole correlation is 
maximum when the hole is just one rung away from the impurity. This is 
expected from  figure 3 for the present choice of parameters ($J=t=t^\prime$).  When the hole is one rung away 
from the impurity, all other parameters remaining the same, 
the magnitude of the correlation function is maximum for AFM impurity and minimum 
for FM impurity. Actually For AFM impurity all the three basis states 
$\phi(1)$, $\theta(1)$ and $\psi(1)$ (all of them have the hole occupying the first rung) contributes to a large extent to the correlation function. For FM impurity there is little contribution form the basis state $\psi(1)$ due to unfavourable configuration (two of the three bonds are frustrated). However, if the hole and impurity are on the same rung, the correlation function has largest value for NM impurity and smallest value for AFM impurity.

\begin{figure}[H]
\centering
\includegraphics[width=11cm,height=6.5cm]{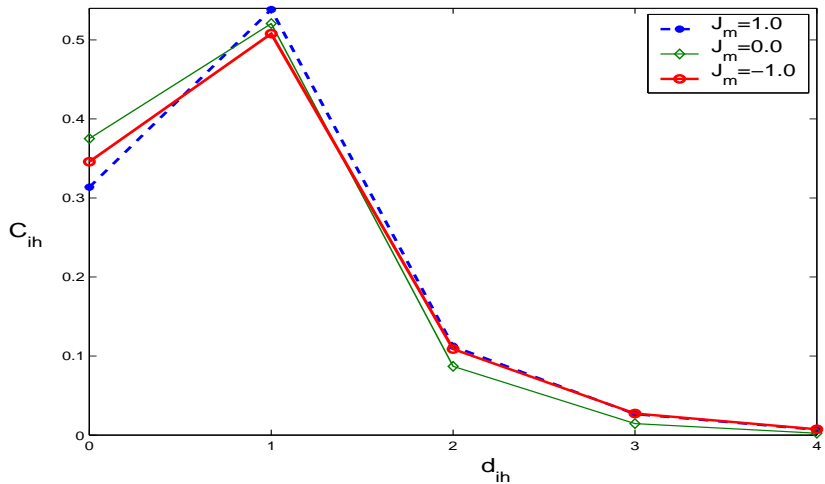}
\caption{Impurity-hole correlation $C_{ih}$ in the ground state as a function 
of the distance $d_{ih}$ of the hole from the impurity (FM, AFM and NM)
($J=t=t^\prime=1.0$, $N=50$).}
\end{figure}

\section*{5. Conclusions}
In this paper we have solved analytically the many-body problem of a single 
hole in presence of a NM impurity and infinite number of electrons for a 
ladder of infinite length. A continuum of scattering states and antibound 
state are obtained in the triplet sector. In the singlet sector, in addition to the 
states found in the triplet sector, bound states of the hole with the impurity 
are also found. As the ratio $J/t$ is increased from zero, successive bound 
states in the symmetric, antisymmetric and symmetric channel appear. For large 
values of $J/t$ all the three bound states co-exist. As pointed out by 
Poilblanc $et al.$ \cite{Poilblanc1} this is a characteristic feature of 
strong correlation. Our analytical calculations for the ladder model can 
qualitatively confirm the numerical results for small 2D clusters by Poilblanc 
$et al.$ \cite{Poilblanc1, Poilblanc2}. However, contrary to the observation by 
Poilblanc $et al.$ we find that the binding energy curve for the impurity-hole 
bound state (in the symmetric sector) is lower than that of a bound pair of 
holes in almost the whole of the parameter space and this definitely indicates 
a strong pair breaking effect \cite{Gayen2} by NM impurity in the 
$t-J$ ladder model. In the case of single hole in presence of a magnetic impurity, the eigenvalue problem is solved numerically for a finite-sized ladder.
Since the number of basis states increases as $N$ we can solve a large size 
ladder with relative ease. Again, the results obtained are in qualitative
 agreement with those obtained by Poilblanc $et al.$ \cite{Poilblanc2} for 2D 
finite clusters. Impurity-hole binding is more favourable in the case of FM 
impurity than in the case of AFM inpurity. With increase in the strength of the
 AFM interaction of the impurity with surrounding electronic spins, the 
magnitude of binding energy rapidly decreases to zero. Comparison
 of the results between
magnetic and NM impurites indicates that NM impurities should act as stronger 
pair-breaking agent in strongly correlated systems than magnetic impurities.

\end{document}